\documentstyle[prl,aps,epsf]{revtex}
\begin{document}
\draft
\twocolumn[
\hsize\textwidth\columnwidth\hsize\csname@twocolumnfalse\endcsname
\preprint{}
\title{
Quasiparticle States at a $d$-Wave Vortex Core in High-$T_c$
Superconductors: Induction of Local Spin Density Wave Order
}
\author{Jian-Xin Zhu and C. S. Ting}
\address{Texas Center for Superconductivity and Department of Physics, 
University of Houston, Houston, TX 77204
}
\maketitle
\begin{abstract}
{ 

The local  density of states (LDOS) at one of the vortex lattice
cores in a high Tc superconductor is studied by using a self-consistent 
mean field theory including interactions for both antiferromagnetism (AF) 
and $d$-wave superconductivity (DSC). The parameters are chosen in such a 
way that in an optimally doped sample the AF order is completely 
suppressed while DSC prevails. In the mixed state, we show that the local 
AF-like SDW order  appears near the vortex core and acts as an
effective local magnetic field on the quasiparticles. As a result, the 
LDOS at the core exhibits a double-peak structure near the Fermi level 
that is in good agreement with the STM observations on YBCO and BSCCO. The
presence of local AF order near the votex core is also consistent with
the recent neutron scattering experiment on LSCO.

}
\end{abstract} 
]

\narrowtext

The quasiparticle states at the vortex core in the mixed state
of a superconductor have been one of the major interest in condensed
matter physics. For an $s$-wave superconductor, the energy gap opened
at the Fermi  surface is a constant and it was predicted long time ago
by Caroli et al.~\cite{Caro64} 
that there should exist the low-lying bound quasiparticle states
inside an $s$-wave vortex core. This prediction was later confirmed by
detailed numerical computations~\cite{Gygi90,Shore89,Zhu95} 
and by STM experiments on NbSe$_2$~\cite{Hess89}
 although
 a direct observation of the
discrete levels is yet to be performed. However,
for a $d$-wave pairing
state as recently established in high-$T_c$ cuprates,
the situation becomes more complex, mostly due to the fact that the energy
gap is closed at the nodal direction on the essentially cylindrical
Fermi surface. In an earlier study by Wang and MacDonald based on a lattice
model~\cite{Wang95}, it
was shown that the local density of states (LDOS) at the $d$-wave vortex core
exhibits a single broad peak at zero energy. Recent low temperature
STM experiments on YBa$_2$Cu$_3$O$_{7-\delta}$ (YBCO)~\cite{Maggio95}(a) 
and
Bi$_2$Sr$_2$CaCu$_2$O$_{8+\delta}$ (BSCCO)~\cite{Pan00} both observed a 
double-peak structure around zero bias
in the local differential tunneling conductance at the vortex core center,
which identifies the widely split core states at energies $\pm 5.5$ meV
and $\pm 7$ meV, respectively. The discrepancy between the theory and the
experiment stimulated further theoretical 
studies~\cite{Himeda97,Morita97,Franz98,Taki99,Yasui99}
on the quasiparticles in the vortex core of high-$T_c$ cuprates.
Franz and Te\v{s}onovi\'{c}~\cite{Franz98} proposed an explanation of the 
observed 
double-peak structure in terms of a mixed $d_{x^2+y^2}+id_{xy}$ pairing 
state. Such a pairing state was previously suggested~\cite{Laugh98} to be 
realized through the field-induced second phase transition 
as motivated by the observation of a plateau in thermal 
conductivity~\cite{Krish97}. The origin of this plateau is still 
hotly debated~\cite{Aubin99}. By pointing out that the parameter 
values are unrealistically chosen in Ref.~\cite{Franz98} 
that the $d_{x^2+y^2}+id_{xy}$ state even 
already exists in zero magnetic field, Yasui and Kita~\cite{Yasui99} 
applied the  ``Landau-level expansion method'' to study the quasiparticles 
in $d$-wave vortex lattice based on the continuum model. They found that the 
double-peak structure may be considered as inherent to systems with 
short coherence length. Since the splitting sensitively depends on 
the field strength, the validity of this scenario 
still needs to be clarified theoretically and experimentally. 
On the other hand, the 
superconducting vortex with antiferromagnetic (AF) core 
was also predicted based on either the SO(5) theory~\cite{Arovas97} or 
the standard $t$-$J$ model with spin-charge separation~\cite{Lee00},
which leads to a featureless LDOS and is 
in disagreement with the STM data 
on the optimally doped YBCO~\cite{Maggio95}(a) and BSCCO~\cite{Pan00}.    
Partly motivated by the observation of magnetic vortex cores in recent 
neutron scattering experiment by Lake {\em et al.}~\cite{Lake01} on optimally
doped
La$_{2-x}$Sr$_x$CuO$_4$, we present in this Letter an alternative 
mechanism for the double-peak structure in the LDOS around zero energy. 
We show that due to electron correlations, the AF-like spin density wave 
(SDW) 
order can develop locally around the vortex core and vanish in the 
superconducting regions. The lift of the spin degeneracy leads to 
the splitting of the zero energy peak. 
The induced SDW order around the vortex core manifests the 
repulsive electron interaction responsible for the strong spin 
fluctuations in the underdoped region of high-$T_c$ cuprates.   
In fact, the coexistence of superconducting (SC) 
and SDW orders has been theoretically  
studied~\cite{Bale98,Sach99,Han00,Mart00}, 
which shows a rich phase diagram with a classic AF order at half filling, 
striped phase at underdoping, a $d$-wave SC at optimal doping.

We start with a generalized Hubbard model
defined on two dimensional (2D) lattice.
By assuming that the on-site repulsion is solely responsible for the
antiferromagnetism while the nearest neighbor attraction causes
the $d$-wave superconductivity, we can construct an effective mean-field
(MF) model~\cite{Mart00}
to study the vortex physics in the
mixed state:
\begin{eqnarray}
H&=&-t\sum_{{\bf ij},\sigma} c_{{\bf i}\sigma}^{\dagger}c_{{\bf j}\sigma}
e^{i \varphi_{\bf ij}}
+\sum_{{\bf i},\sigma} (m_{{\bf i},\bar{\sigma}}-\mu)
c_{{\bf i}\sigma}^{\dagger}c_{{\bf i}\sigma} \nonumber \\
&&+\sum_{\bf ij} (\Delta_{\bf ij} c_{{\bf i}\uparrow}^{\dagger}
c_{{\bf j}\downarrow}^{\dagger}
+\Delta_{\bf ij}^{*} c_{{\bf j}\downarrow} c_{{\bf i}\uparrow} ) \;.
\label{EQ:MFA}
\end{eqnarray}
Here $c_{{\bf i}\sigma}$ annihilates an electron of
spin $\sigma$ at site ${\bf i}$.
The summation is over the nearest neighbor sites.
$\mu$ is the chemical potential.
$m_{{\bf i},\sigma}=U \langle c_{{\bf i}\sigma}^{\dagger} 
c_{{\bf i}\sigma}\rangle$ is
the spin-dependent Hartree-Fock potential at site ${\bf i}$, where $U$ is 
the strength of on-site
repulsion.
$\Delta_{\bf ij}=\frac{V}{2}\langle c_{{\bf i}\uparrow}
c_{{\bf j}\downarrow} -c_{{\bf i}\downarrow}c_{{\bf j}\uparrow}\rangle$
is the spin-singlet $d$-wave pair potential, where $V$ is the strength of
nearest neighbor effective electron-electron attraction.
As a phenomenological model, we do not intend to address the microscopic
mechanism for this attraction.
In the mixed state, the magnetic field effect was included through the
Peierls phase factor $\varphi_{\bf ij}=\frac{2\pi}{\Phi_{0}}
\int_{{\bf r}_{\bf j}}^{{\bf r}_{\bf i}} {\bf A}({\bf r})\cdot d{\bf r}$,
where $\Phi_0=hc/2e$ is the superconducting flux quantum.
By assuming the superconductor under consideration is in the extreme type-II
limit where the Ginzburg-Landau parameter $\kappa=\lambda/\xi$ goes to
infinity so that the screening effect from the supercurrent is negligible. 
Therefore, the vector potential ${\bf A}$ can be approximated by the
solution $\nabla \times {\bf A}=H \hat{\bf z}$ where $H$ is the magnetic
field externally applied along the $c$ axis. The enclosed flux density
within each plaquette is given by $\sum_{\Box} \varphi_{\bf ij}
=\frac{2\pi Ha^{2}}{\Phi_0}$. A similar mean-field Hamiltonian can
also be arrived at within a $t$-$U$-$J$ model  proposed 
recently~\cite{Daul00}.
 
We diagonalize the Hamiltonian Eq.~(\ref{EQ:MFA}) by solving the
BdG equation:
\begin{equation}
\sum_{\bf j} \left(
\begin{array}{cc}
{\cal H}_{{\bf ij},\sigma} & \Delta_{\bf ij}  \\
\Delta_{\bf ij}^{*} & -{\cal H}_{{\bf ij},\bar{\sigma}}^{*}
\end{array}
\right) \left(
\begin{array}{c}
u_{{\bf j}\sigma}^{n} \\ v_{{\bf j}\bar{\sigma}}^{n}
\end{array}
\right)
=E_{n}
\left(
\begin{array}{c}
u_{{\bf i}\sigma}^{n} \\ v_{{\bf i}\bar{\sigma}}^{n}
\end{array}
\right)  \;,
\label{EQ:BdG}
\end{equation}
where $(u_{{\bf i}\sigma}^{n},v_{{\bf i}\bar{\sigma}}^{n})$
is the quasiparticle wavefunction
corresponding to the eigenvalue $E_n$, the single particle Hamiltonian
${\cal H}_{{\bf ij},\sigma}=-t e^{i\varphi_{\bf ij}} \delta_{
{\bf i}+\boldmath{\mbox{$\delta$}},{\bf j}} +
(m_{{\bf i},\bar{\sigma}}-\mu)\delta_{\bf ij}$. Notice that the quasiparticle
energy is measured with respect to the Fermi energy.
The self-consistent conditions read:
\begin{equation}
m_{{\bf i}\sigma}=U\sum_{n} \vert u_{{\bf i}\sigma}^{n}\vert^{2} f(E_n)\;,
\end{equation}
and
\begin{equation}
\Delta_{\bf ij}=\frac{V}{4}\sum_{n}
(u_{{\bf i}\uparrow}^{n}v_{{\bf j}\downarrow}^{n*}
+v_{{\bf i}\downarrow}^{n*}u_{{\bf j}\uparrow}^{n}
)
\tanh \left( \frac{E_{n}}{2k_{B}T}\right)\;,
\end{equation}
where the Fermi distribution function $f(E)=1/(e^{E/k_{B}T}+1)$.
Here the summation is also over those eigenstates with negative eigenvalues 
thanks to the symmetry property of the BdG equation:  
If $(u_{{\bf i}\uparrow}^{n},u_{{\bf i}\downarrow}^{n},
v_{{\bf i}\uparrow}^{n},
v_{{\bf i}\downarrow}^{n})^{Transpose}$ is the eigenfunction of the 
$4\times 4$
equation in the spin space with energy $E_n$, then
$({v_{{\bf i}\uparrow}^{n}}^{*},-{v_{{\bf i}\downarrow}^{n}}^{*},
{u_{{\bf i}\uparrow}^{n}}^{*},
-{u_{{\bf i}\downarrow}^{n}}^{*})^{Transpose}$ up to a global phase 
factor  
is the eigenfunction with energy $-E_n$.

Hereafter we measure the length in units of the
lattice constant $a$ and the energy in units of the hopping integral $t$.
Within the Landau gauge the vector potential can be written as
${\bf A}=(-H y,0,0)$ where $y$ is the $y$-component of the position
vector {\bf r}.
We introduce the magnetic translation operator
${\cal T}_{mn}{\bf r}={\bf r}+{\bf R}$
where the translation vector ${\bf R}=m N_x \hat{\bf e}_{x} +n N_y
\hat{\bf e}_{y}$ with $N_{x}$ and $N_{y}$ the linear dimension of the unit 
cell of the vortex lattice.
To ensure different ${\cal T}_{mn}$ to be commutable with each other,
we have to take the strength of magnetic field so that the flux enclosed
by each unit cell has a single-particle flux quantum, i.e, $2\Phi_0$.
Therefore, the translation property of the superconducting order parameter
is $\Delta ({\cal T}_{mn}{\bf r})=e^{i\chi({\bf r},{\bf R})} \Delta({\bf 
r})$ where the phase accumulated by the order parameter upon the 
translation   is
$\chi({\bf r},{\bf R})=\frac{2\pi}{\Phi_0}{\bf A}({\bf R}) \cdot {\bf r}
-4mn\pi$. From this property, we can obtain the magnetic Bloch theorem for
the wavefunction of the BdG equations:
\begin{equation}
\left(
\begin{array}{c}
u_{{\bf k},\sigma}({\cal T}_{mn}\tilde{\bf r}) \\
v_{{\bf k},\sigma}({\cal T}_{mn}\tilde{\bf r})
\end{array}
\right)
 = e^{i{\bf k}\cdot {\bf R}}
\left(
\begin{array}{c}
e^{i\chi({\bf r},{\bf R})/2} u_{{\bf k},\sigma}(\tilde{\bf r}) \\
e^{-i\chi({\bf r},{\bf R})/2}v_{{\bf k},\sigma}(\tilde{\bf r})
\end{array}
\right) \;.
\end{equation}
Here $\tilde{\bf r}$ is the position vector defined within a given unit
cell and ${\bf k}=\frac{2\pi l_x}{M_x N_x}\hat{\bf e}_{x} +
\frac{2\pi l_y}{M_y N_y}\hat{\bf e}_{y}$ with 
$m_{x,y}=0,1,\dots,M_{x,y}-1$
are the wavevectors defined in the first Brillouin zone of the vortex 
lattice and $M_x N_x$ and  $M_y N_y$ are  the linear dimension of the whole 
system. The vortex carrying the flux quantum $hc/2e$ is the generic 
feature of
the pairing theory for superconductivity. Therefore, it is not surprising
that in the slave boson approach to the $t$-$J$ model, the vortex always
carries $hc/2e$ flux quantum if the magnetic 
field is assumed~\cite{Lee00} 
to act on electrons only through the spinon degrees of freedom.

As a model calculation, we take the following parameter values: The 
pairing interaction is $V=1.0$, and the filling factor, which is
defined as $n_f=\sum_{{\bf i},\sigma} \langle
c_{{\bf i}\sigma}^{\dagger}c_{{\bf i}\sigma}\rangle /N_x N_y$ with the 
summation over one unit cell, is fixed to be $0.84$ so that
the chemical potential needs to be adjusted each time
the on-site repulsion $U$ is varied. For our interest in the low energy 
quasiparticle states, we only consider the zero temperature limit.
We have typically considered the 
unit cell of size $N_x\times N_y= 42\times 21$,
and the number of the unit cells
$M_x \times M_y=21\times 42$. This choice will give us a square vortex 
lattice. We use exact diagonalization method to
solve the BdG equation~(\ref{EQ:BdG}) self-consistently: To allow the
inhomogeneity of all physical quantities, randomly distributed
$\Delta_{\bf ij}$ and $m_{{\bf i},\sigma}$ are taken as initial 
parameters; the newly obtained $\Delta_{\bf ij}$ and $m_{{\bf i},\sigma}$ 
are then substituted back into the equation. The above procedure is 
repeated until the convergence with required accuacy is achieved.
In the absence of magnetic field, we have reproduced the results 
reported in previous work~\cite{Mart00} including an AF SDW 
order, a stripe phase, and a $d$-wave SC phase 
when the system is doped away from the undoped to the optimally doped 
region. In the present work, we are mainly concerned with 
the electronic structure around the vortex core in the optimally doped 
region. In this region, the SDW order is strongly suppressed and the 
$d$-wave SC order is homogeneous in real space. 
However, when a magnetic field is applied to drive the system into the
mixed state so that the $d$-wave order parameter is suppressed around the 
vortex core, we find that as the on-site repulsion is increased to about 
$1.5$, the SDW order is nucleated around the vortex core. Typical results 
on the nature of the vortex core is displayed in Fig.~\ref{FIG:VORTEX} 
with the on-site repulsion $U=2$. As shown in Fig.~\ref{FIG:VORTEX}(a), 
each unit cell accommodates two superconducting vortices each carring a 
flux quantum $hc/2e$. The $d$-wave SC order parameter vanishes at the 
vortex core center and starts to increase at the scale of 
the coherence length $\xi_0$ to its bulk value
which is about 0.1 for the chosen parameter values. 
Fig.~\ref{FIG:VORTEX}(b) displays the spatial distribution of the 
staggered magnetization of the local SDW order as defined by 
$M_{s}=(-1)^{\bf i}S_{z}^{\bf i}$ with
$S_{z}^{\bf i}=n_{{\bf i},\uparrow}-n_{{\bf i},\downarrow}$.
Clearly, the maxima strength
of $M_{s}$ appears at the vortex core center and decays also with a scale 
of $\xi_0$ to zero into the superconducting region.
More interestingly, the SDW order parameters has opposite
polarity around two nearest neighbor vortices along the $x$ 
direction.  We have compared the free energy between this 
configuration with that obtained by switching the  orientation of the SDW 
order around one of the two nearest neighbor vortices, and found that the 
free energy for both configurations is very close but it is always lower 
in the former case. 
Therefore, the induction of the SDW order around the 
vortices reduces the four-fold rotational symmetry of the whole system to 
the two-fold, and the period of the translational symmetry of the vortex 
lattice along the $x$ is doubled. This result is  
understandable when we notice 
the zero-field result~\cite{Mart00}: 
The homogeneous superconducting order in the 
optimally doped region derives from the melting of the strongly 
overlapped quasi-one dimensional (Q1D) superconducting stripes (i.e, 
soliton-like AF anti-phase domain boundaries, at which 
the AF SDW order changes sign). The development of the Q1D 
stripes breaks at the beginning the four-fold rotational 
symmetry of 
the 
system. Therefore, it seems that the development of the local SDW order 
around the vortices in the optimally doped region can be regarded as
a duality of the development of the local SC order around the AF stripe 
in the underdoped region. On the other hand, 
the appearance of the SDW order around the vortex strongly affects the 
electron density $n_{\bf i}=\sum_{\sigma} n_{{\bf i},\sigma}$. As shown in 
Fig.~\ref{FIG:VORTEX}(c), at the vortex core center, where the SDW 
amplitude 
reaches the maximum, the electron density is strongly enhanced and is 
very close to unity, which is characteristic of the bulk AF-like SDW 
order at the half filling. Therefore, the hole charge density is depleted 
in the vortex core center. The depletion of the hole charges 
near the vortex core center is compensated by the corresponding 
enhancement  
along $\pi/4$ and $3\pi/4$ directions with respect to the underlying
crystal lattice, which correspond to the nodal directions on
 the Fermi
surface.
This kind of charge
inhomogeneity is closely related to the development of the local SDW order     
around the vortex core due to the large on-site repulsion.
Finally, with the chosen parameter values, we also find that
when the on-site repulsion $U$ is increased to $3.5$, the SC vortex 
state is collapsed and the SDW order becomes dominant. Our numerical 
analysis seems to be consistent with the recent argument~\cite{Demler01} 
of the magnetic field driven quantum phase transition from the SC state 
into a state with microscopic coexistence of SC and SDW orders to 
understand the recent neutron scattering experiments~\cite{Lake01}.
Notice that the quantum fluctuation of the SDW order parameter, which 
plays an important role in addressing correctly the spin excitations, has 
been neglected in the present model because our interest is in the 
quasiparticle states. 

We now turn our attention to the quasiparticle state at the vortex 
core center. The local density of states is defined by  
\begin{equation}
\rho_{\bf i}(E)=-\frac{1}{M_x M_y} \sum_{{\bf k},n,\alpha} 
\vert u_{{\bf k},{\bf i},\sigma} \vert^{2} f^{\prime}(E^{n}_{\bf k}-E)
\;,
\end{equation}  
where $f^{\prime}(E)$ is the derivative of the Fermi distribution 
function. $\rho_{\bf i}(E)$ is
proportional to the local differential tunneling conductance which could 
be   
measured by STM
experiments~\cite{Tinkham75}. 
In Fig.~\ref{FIG:LDOS} we plot the LDOS as a function
of energy at the vortex core center for different values of 
on-site repulsion. 
For comparison, we have also displayed the LDOS at the 
midpoint between two nearest-neighbor vortices along the $x$ direction, 
which resembles that for the bulk system. 
The asymmetry line shape in $\rho_{\bf i}(E)$ with respect
to zero energy reflects
the lack of particle-hole symmetry as the chemical potential $\mu$ 
deviates from zero for $n_f$ being less than the half filling ($n_f=1$). 
As can be seen from Fig.~\ref{FIG:LDOS}(a), when $U=0$ for which no local
SDW order is induced,
the LDOS at the core center shows a single resonant peak around 
the Fermi energy, which is similar to that reported by other
authors~\cite{Wang95}.
When $U$ is sufficiently large that the local SDW order develops around 
the vortex core, the LDOS peak at zero energy is split into a double-peak
structure 
(see Fig.~\ref{FIG:LDOS}(b)). The splitting comes from the fact: When   
 the SDW order is localized around the vortex center, 
the spin-dependent potential, which can be rewritten as 
$m_{{\bf i},\uparrow(\downarrow)}=U(n_{\bf i}\pm S_{z}^{\bf i})/2$,
plays the role of a local magnetic field  interacting with the electrons 
via the Zeeman coupling. When $U$ is increased, $S_{z}^{\bf i}$ at
the vortex core center is enhanced,
and the combination of them  enlarges the Zeeman
interaction. As a consequence, the LDOS peak at zero energy is further 
split (see Fig.~\ref{FIG:LDOS}(c)). This splitting of the LDOS at the 
vortex core center is in good agreement with the STM experiments on the 
optimally doped YBCO and BSCCO. The induction of the local AF order at the 
vortex core is consistent with recent neutron scattering experiment on 
LSCO~\cite{Lake01}.

{\bf Acknowledgments}: We wish to thank A.V. Balatsky, E. Demler, B. 
Friedman, T. Kita, T.K. Lee, A.H. MacDonald, S. Sachdev, D.N. Sheng, 
W.P. Su, M. Takigawa, K.K. Voo, and Z.Y.  Weng for useful discussions. 
This work was 
supported by the Texas Center  for Superconductivity at the University of 
Houston through the State of
Texas, and the Robert A. Welch Foundation, and the ARP-0036520241-1999.

\begin{figure}
\caption[*]{The amplitude distribution of the $d$-wave SC order parameter 
$\vert \Delta_{d}\vert$ (a), the staggered  magnetization 
$M_{s}$  (b), and the electron density $n_{\bf i}$ (c) in 
one  magnetic unit cell obtained at the zero temperature. The size of the 
cell  is $42\times 21$. The strength of the on-site repulsion $U=2$. The 
other parameter values: The $d$-wave pairing interaction $V=1$; the filling 
factor $n_f=0.84$.    
}
\label{FIG:VORTEX}
\end{figure}

\begin{figure}
\caption[*]{
The zero-temperature LDOS at the 
vortex core center (red-solid line) with various strength of on-site 
repulsion $U=0$ (a), $U=2$ (b), and $U=3$ (c). Also displayed is the 
zero-temperature LDOS at the midpoint (green-dot-dashed line) 
between two nearest neighbor 
vortices along the $x$ direction. The other parameter values are the same 
as Fig.~\ref{FIG:VORTEX}.
}
\label{FIG:LDOS}
\end{figure}


\begin{references}
\bibitem{Caro64} C. Caroli, P. G. de Gennes, and J. Matricon, Phys. Lett. 
{\bf 9}, 307 (1964).

\bibitem{Gygi90} F. Gygi and M. Schluter, Phys. Rev. Lett. {\bf 65}, 1820 
(1990); Phys. Rev. B {\bf 41}, 822 (1990); {\bf 43}, 7609 (1991). 

\bibitem{Shore89} J. D. Shore {\em et al.}, Phys. Rev. Lett. {\bf 62},  
3089 (1989).

\bibitem{Zhu95} Yu-Dong Zhu, F.C. Zhang, and M. Sigrist, Phys. Rev. B {\bf 
51}, 1105 (1995).

\bibitem{Hess89} H. F. Hess {\em et al.}, Phys. Rev. Lett. {\bf 62}, 214 
(1989).

\bibitem{Wang95} Y. Wang and A.H. MacDonald, Phys. Rev. B {\bf 52}, 
R3876 (1995).

\bibitem{Maggio95} (a) I. Maggio-Aprile {\em et al.}, Phys. Rev. Lett. 
{\bf 75}, 2754 (1995); (b) Ch. Renner {\em et al.}, Phys. Rev.
Lett. {\bf 80}, 3606 (1998).

\bibitem{Pan00} S. H. Pan {\em et al.}, Phys. Rev. Lett. {\bf 85}, 1536 
(2000).

\bibitem{Himeda97} A. Himeda {\em et al.}, J.
Phys. Soc. Jpn. {\bf 66}, 3367 (1997). 

\bibitem{Morita97} Y. Morita, M. Kohmoto, and K. Maki,
Phys. Rev. Lett. {\bf 78}, 4841 (1997); {\bf 79}, 4514 (1997);
M. Franz and M. Ichioka, {\em ibid.} {\bf 79}, 4513 (1997). 

\bibitem{Franz98} M. Franz and Z. Te\v{s}anovi\'{c}, Phys. Rev. Lett. 
{\bf 80}, 4763 (1998).

\bibitem{Taki99} M. Takigawa, M. Ichioka, and K. Machida, Phys. Rev. Lett. 
{\bf 83}, 3057 (1999). 

\bibitem{Yasui99} K. Yasui and T. Kita, Phys. Rev. Lett. {\bf 83}, 
4168 (1999).

\bibitem{Laugh98} R. B. Laughlin, Phys. Rev. Lett. {\bf 80}, 5188 (1998).

\bibitem{Krish97} K. Krishana {\em et al.}, Science {\bf 277}, 83 (1997).

\bibitem{Aubin99} H. Aubin {\em et al.}, Phys. Rev. Lett. {\bf 82}, 624 
(1999).

\bibitem{Arovas97} D. P. Arovas {\em et al.}, Phys. Rev. Lett. {\bf 79}, 
2871 (1997).

\bibitem{Lee00}  Jung Hoon Han and D.-H. Lee, Phys. Rev. Lett. {\bf 85}, 
1100 (2000).

\bibitem{Lake01} B. Lake {\em et al.}, Science {\bf 291}, 1759 (2001); 
cond-mat/0104026.

\bibitem{Bale98} L. Balents, M. P. A. Fisher, and C. Nayak, Int. J. Mod. 
Phys. B {\bf 12}, 1033 (1998).

\bibitem{Sach99} S. Sachdev, C. Buragohain, and M. Vojta, Science {\bf 
286}, 2479 (1999).

\bibitem{Han00} J. H. Han, Q.-H. Wang, and D.-H. Lee, cond-mat/0006046.

\bibitem{Mart00} I. Martin {\em et al.}, cond-mat/0009067.

\bibitem{Daul00} S. Daul, D. J. Scalapino, and S. R. White, Phys. Rev. 
Lett. {\bf 84}, 4188 (2000). 

\bibitem{Demler01} E. Demler, S. Sachdev, and Y. Zhang, cond-mat/0103192.

\bibitem{Tinkham75} M. Tinkham, {\em Introduction to
Superconductivity} (McGraw Hill, New York, 1975).


\end{references}
\end{document}